\documentclass[aps,prl,amsmath,amssymb,twocolumn,nofootinbib]{revtex4}
\usepackage{pslatex,graphicx,dcolumn,bm,natbib}
\date{\today}

\begin{document}


\title{Why Do Granular Materials Stiffen with Shear Rate?
A Test of  Novel Stress-Based Statistics}

\author{R.P. Behringer$^1$, Dapeng Bi$^2$, B. Chakraborty$^2$, S. Henkes$^2$, and
  R. R. Hartley$^1$\\
\normalsize{$^1$Department of Physics, Duke University, Box 90305,
Durham, NC 27708, USA\\$^2$Department of Physics, Brandeis University,
Waltham, MA 02454, USA} }

\begin{abstract}

Recent experiments exhibit a rate-dependence for granular shear such
that the stress grows linearly in the logarithm of the shear rate,
$\dot{\gamma}$.  Assuming a generalized activated process mechanism,
we show that these observations are consistent with a recent proposal
for a stress-based statistical ensemble.  By contrast, predictions for
rate-dependence using conventional energy-based statistical mechanics
to describe activated processes, predicts a rate dependence that of $(\ln (\dot{\gamma}))^{1/2}$.

\end{abstract}

-
\maketitle

Understanding disordered solids, such as foams, glasses, polymers,
colloids and granular materials is a great challenge for statistical
physics.  Several of these systems, including granular materials, fall
outside the rubric of conventional statistical mechanics because they
are dissipative.
But, these materials exhibit well defined statistical distributions.  Several novel approaches have
been recently
proposed\cite{edwards89,makse_02,ono_02,ohern_04,snoeijer2,tighe,henkes05,henkes06} to characterize the statistics of these dissipative materials,.
We focus on testing one of the proposed statistical frameworks, the force or stress-based
ensembles\cite{snoeijer2,tighe,henkes05,henkes06}, and specifically the 
stressed-based ensemble hypothesized by SH and BC\cite{henkes05,henkes06} to account for the coupling between
forces and geometry.  Here, we test this hypothesis by showing that it
can account for experimentally observed logarithmic strengthening with
increasing shear rate in slowly sheared granular systems.  Many models
of slow, dense granular flows assume that the internal stresses are
independent of shear rate.  Linear rate-dependence for shear stresses
occurs for Newtonian fluids. Non-linear dependence on rate is common
in non-Newtonian fluids, as well as in the ``glassy'' systems noted
above\cite{sollich_prl97,sollich98}. The stress-based ensemble offers
an explicit framework for analyzing the rheology of such non-thermal,
glassy systems.

\begin{figure}
\includegraphics[width=1.5in]{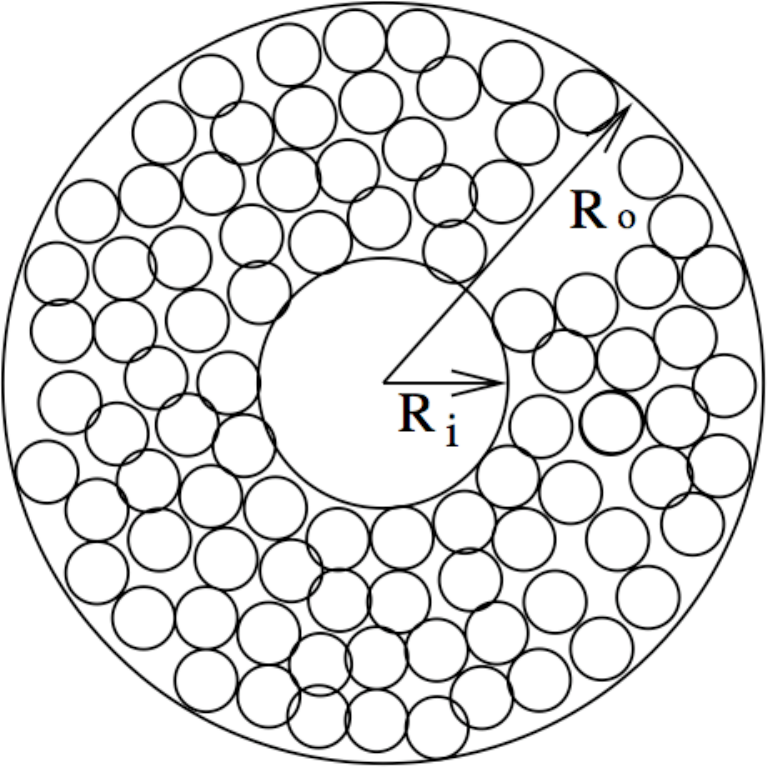}%
\vspace{0.15in}
\newline

\includegraphics[width=2.5in]{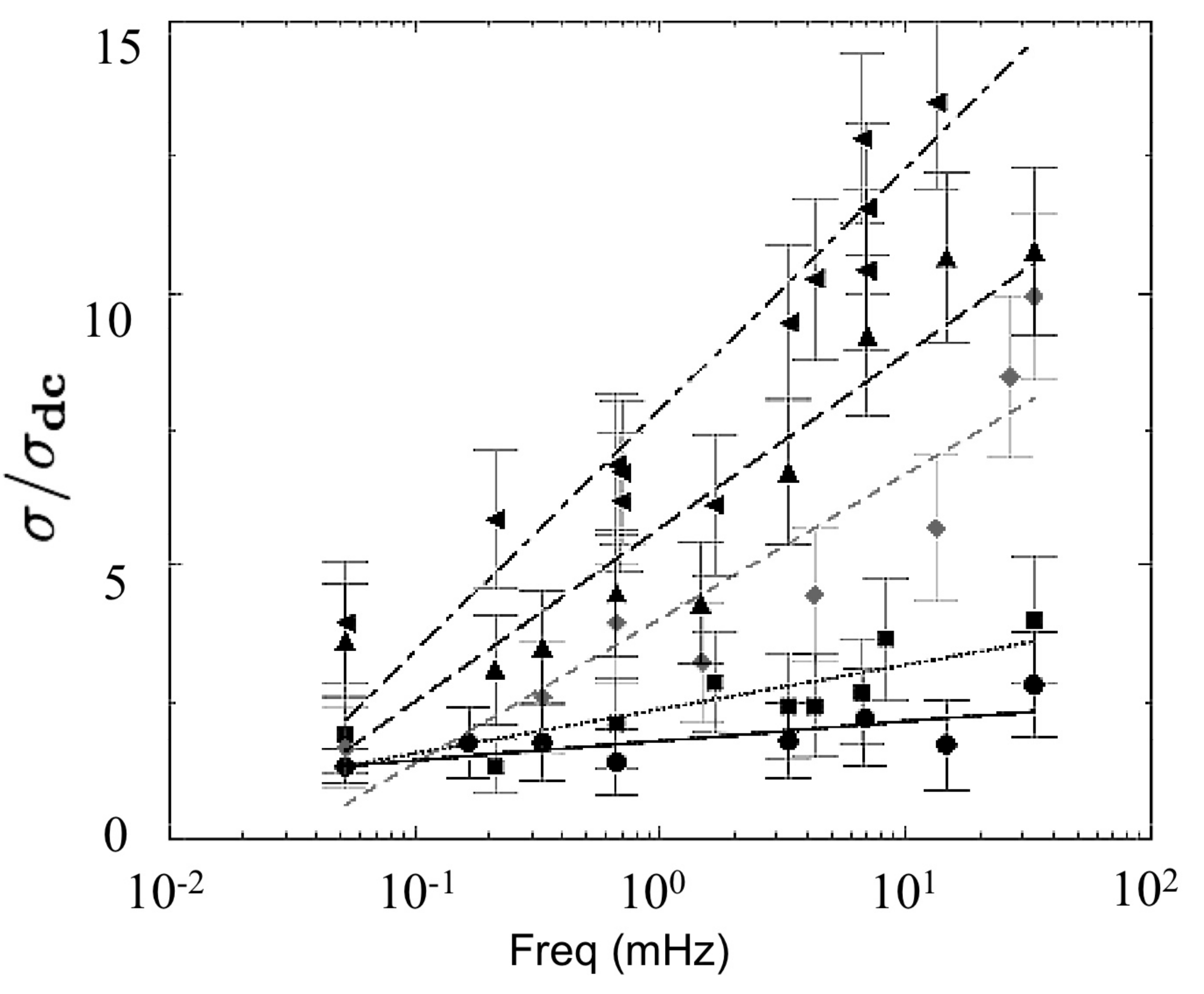}%

\caption{\label{fig:loggam} Top, sketch of Couette shear apparatus,
  not to scale.  $R_o = 19.2\, cm$.  Data is for small/medium shearing
  wheels, where $R_i = 6.7\, cm/ 5.3\, cm$.  Particles are either
  circular or pentagonal in cross section, and have dimensions between
  $0.7\, cm$ and $0.9\, cm$.  Bottom: Data for the mean stress in a
  segment of a 2D granular Couette experiment containing roughly 200
  particles vs. shear $\dot{\gamma}$.  Data (normalized by $\rm
  \sigma_{dc}\simeq 4.11 Nm^{-1}$) are for different packing fractions
  $\phi$ relative to the critical packing fraction, $\phi_c$, where
  stesses fall to zero.  Symbols are: circles: $\phi - \phi_c =
  0.0012$, squares: $\phi - \phi_c = 0.0091$, diamonds: $\phi - \phi_c
  = 0.0152$, upward pointing triangles: $\phi - \phi_c = 0.0189$,
  left-pointing triangles: circles: $\phi - \phi_c = 0.0226$.  We also
  consider other data (below) from experiments by Hartley and
  Behringer\cite{hartley_03}.  }
\end{figure}

\begin{figure}
\centerline{\includegraphics[width=2.5in]{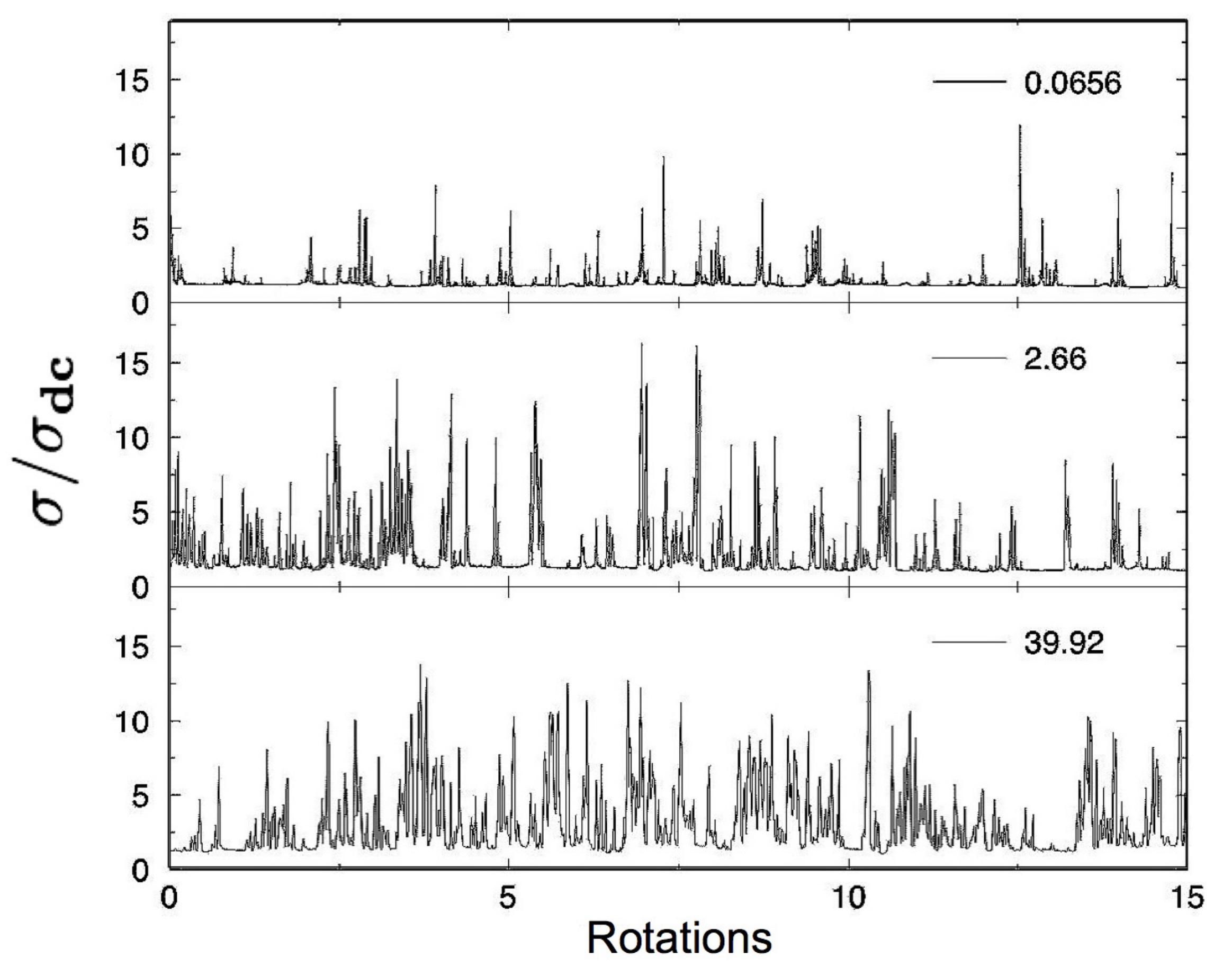}}
\caption{\label{fig:timeser} Time series for different $\dot{\gamma}$
  (in mHz) for the stress, for the smaller $R_i$ above.}
\end{figure}

Recent 2D \cite{hartley_03} and 3D\cite{daniels05a} experiments on
sheared dense granular materials, showed mean stresses that grow
linearly with $\ln (\dot \gamma)$, where $\dot \gamma$ is the shear
rate.  Indeed, rate-dependence spans many decades in $\dot \gamma$, as
seen in Fig.~\ref{fig:loggam}, which show time-averaged stresses
acquired in a 2D Couette shear experiment that is sketched in
Fig.~\ref{fig:loggam}.  Fig.~\ref{fig:timeser} shows typical traces of
stress vs. time for several different shear rates, $\dot \gamma$.
These data have also been acquired for various packing fractions,
$\phi$, relatively near the critical packing fraction, $\phi_c$, below
which the system is unjammed\cite{OHern2002,howell_99} and shearing
stops.

\begin{figure}
\centerline{\includegraphics[width=3.0in]{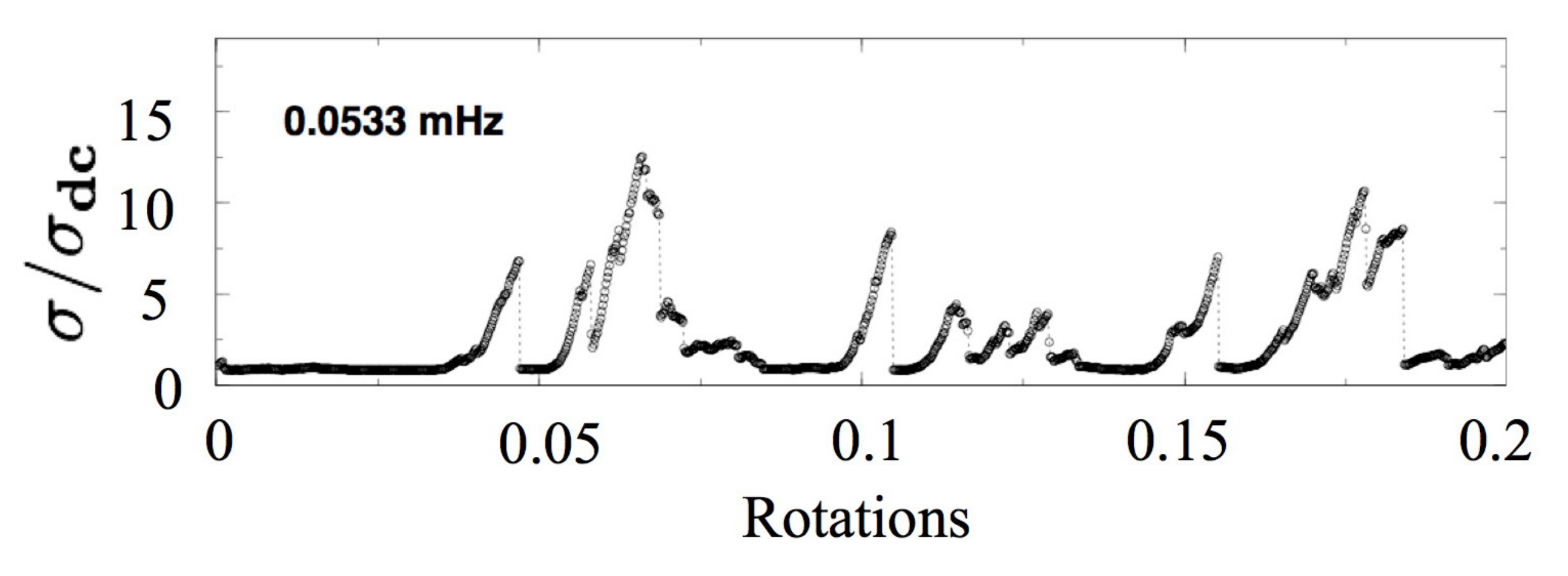}}
\caption{\label{fig:blowup} Blow up of time series data for the stress
  at $\dot{\gamma} = 0.0533mHz$ showing stress fluctuations over 0.2
  rotations of the inner shearing wheel.  The data indicate the
  approximately linear loading up of force structures and their more
  rapid failure.}
\end{figure}

We construct a model for the behavior of the force fluctuations seen
in the Couette shear experiments.  The first model premise is that the
shearing process causes build-up of inhomogeneous stress structures,
such as force chains, which fail when they reach a critical yield
stress.  A blow-up of a typical time series for the stress in
Fig.~\ref{fig:blowup} gives a good flavor of this process.

The actual failure process is complex except very near jamming ($\phi
\simeq \phi_{c}$), where there are typically only one or two visible
force chains in an observation region.  More generally, for $\phi$
much larger than $\phi_{c}$, there is a (strong) force network, and
when a stress drop occurs, it is typically concentrated on a subset of
grains on a more or less linear region of the force network, often
extending for many grains.  The stress drop is very fast relative to
the build-up, indicating a failure of part of the network, while much
of the remaining network is only weakly affected.  We use the common
term ``force chains'' to refer to segments of the strong force network
that exhibit the force build-up and failure process that is the
fundamental origin of granular force fluctuations, but we do not
consider the specific origins of the failures (e.g.
shear-transformation-zone events\cite{lemaitre02}, force chain
buckling\cite{tordesillas08}, etc.).  Rather, we explore the role
played by the background of force fluctuations generated during the
build-up and failure process.  We ask whether the effect of these
fluctuations can be described in terms of a stress-based ensemble.
And we exploit the fact that force changes following failure are
localized to force chains, with a much weaker effect on the rest of
the system.

We expect that a chain will fail if the force/stress on it, $\sigma$,
exceeds a characteristic value $\sigma_m$.  This is reminiscent of
Coulomb failure, but refers to failures of localized structures, with
no strict frictional analogy. We begin
by considering a single event consisting of the birth-to-death cycle
of a single force chain, relevant to systems with $\phi \simeq \phi_{c}$, and  then return to an accounting for multiple
events occuring in a given observation region for packings  with $\phi >> \phi_{c}$.

A key premise of the model is that the failure is an activated process
aided by stress fluctuations, similar to a thermally activated escape
from a potential well.  The potential well is replaced by a stress
trap that models the meso-scale strong force network,  and
  chain failures correspond to escape from a trap.  The fluctuations
of thermal equilibrium are replaced by fluctuations of stress in the
network, characterized by a temperature-like quantity\cite{henkes06}.
These fluctuations occur as the granular assembly moves through a
series of states at mechanical equilibrium.  For thermally activated
processes, the rate of escape is proportional to $e^{-\beta
  E_{barrier}}$, where $\beta = (k_B T)^{-1}$ is the inverse
temperature and $E_{barrier}$ is the barrier height.  To construct a
framework for activated dynamics in systems where the fluctuations are
athermal stress fluctuations, we appeal to a recently-developed
statistical framework for granular assemblies\cite{henkes05,henkes06}.
In this framework, the ensemble of mechanically stable states is
defined by a Boltzman-like probability distribution:
\begin{equation}
P_{\nu} = (1/Z)e^{-\alpha \Gamma_{\nu}} ~,
\label{Boltzman}
\end{equation}
where $\Gamma_{\nu} = S_{\nu} \sigma_{\nu}$ is an extensive quantity
related to the stress of the configuration $\nu$, and $S_{\nu}$ is the
area occupied by the grains\cite{henkes05,henkes06}.  In
Eq. \ref{Boltzman}, $\alpha$ is the analog of the inverse temperature,
$\beta$, and  characterizes the fluctuations.  In analogy with
thermally activated processes, the
probability per unit time of  chain failure is then given by
\begin{equation}
P_f = A \exp[-\alpha \Gamma_{barrier}].
\label{eq:new}
\end{equation}
Assuming that the area fluctuations are small compared to the stress
fluctuations (the area is fixed in the experiments discussed above),
the effective barrier to be surmounted by a force chain with a
stress $\sigma$ on it becomes
$\Gamma_{{barrier}}=S(\sigma_{m}-\sigma)$, and the failure rate (per
unit time) is:
\begin{equation}
P_f = A \exp[-(\sigma_m - \sigma)/\sigma_o)].
\label{eq:0}
\end{equation}
We have used $\sigma_{0}$ to denote
$1/(\alpha S)$.  


Both $A$, the attempt frequency, and $\sigma_o$ may depend on $\dot
\gamma$, but to lowest order, we will treat these as constants.  We
expect that the stress on a force chain increases linearly in
time until  a chain fails (e.g. Fig.~\ref{fig:blowup}).  Thus,
\begin{equation}
\sigma = \sigma' t = {\Sigma}{\dot \gamma} t,
\label{linear}
\end{equation}
where ${\Sigma}$ is a measure of rate of stress increase per unit
shear deformation, which we also assume is a constant.  With this
picture, the  force chain loads up steadily in time, but the
probability of failure depends on the closeness of $\sigma$ to
$\sigma_m$.  If $\sigma << \sigma_m$, the probability of
failure/unit-time should be low, but as $\sigma$ approaches
$\sigma_m$, the probability of failure should become large.  A range
of parameters where the process is strongly activated is $A << \dot
\gamma$, and $\sigma_m >> \sigma_o$, the analog of the low-temperature
limit of a thermally activated process.  This limit likely applies to
the experiments since the time-dependence of the stress is dominated
by slow build up and rapid release.  This is the regime we focus on in
this work.  The assumption of strongly activated behavior is also
consistent with the fact that the experiments suggest logarithmic rate
dependence over many decades of $\dot{\gamma}$, without a crossover to
different behavior at low $\dot{\gamma}$.

We treat the loading up of the networks in a probabilistic
manner.  If the probability for the
chain to survive until time $t$ without failing is $P_s(t)$, then
\begin{equation}
P_s(t + dt) = P_s(t)(1 - P_f(t)dt),
\end{equation}
or
\begin{equation}
P_s^{-1}dP_s/dt = -P_f.
\label{eq:1}
\end{equation}
We first consider the idealized situation of  isolated chains, so
  that we need only consider one such chain at any given instant, and
focus on the evolution of stresses from the formation of the 
  chain to its failure.  The time average moments of $\sigma$ are:
\begin{eqnarray}
\bar{\sigma^{n}}  &=& (1/T) \int_{0}^{T} dt (\sigma (t))^n \nonumber \\
& = & \int  d\tau  P_s(\tau) P_f(\tau)\int_{0}^{\tau} dt (\sigma(t))^n / \int  d\tau  P_s(\tau) P_f(\tau) \tau
\label{eq:2}
\end{eqnarray}
where we have used, $T = \int d\tau P_s(\tau) P_f(\tau) \tau$.
Defining $\langle \sigma^{n} \rangle = \int d\tau P_s(\tau) P_f(\tau)
(\sigma(\tau))^{n}$, we can write $\bar \sigma^{n} = \langle
\sigma^{n+1} \rangle/((n+1)\langle \sigma \rangle )$ The above
equation reflects an ensemble average over chains surviving up to time
$\tau$ and failing within $\tau$ to $\tau + d\tau$, with probability
$P_s(\tau) P_f(\tau)$.
We can use Eq.~\ref{eq:1} to write:
\begin{equation}
\langle \sigma^n \rangle = - \int (\sigma (\tau))^n (dP_s/d\tau) d\tau
\label{eq:8}
\end{equation}
Introducing the dimensionless time,
$\theta = \dot \gamma \tau$, we can write
\begin{equation}
d \ln (P_s)/d \theta = -B \exp ({\Sigma} \theta/\sigma_o),
\label{eq:dlnpdtheta}
\end{equation}
\noindent
where we we have absorbed some of the constants into a single
expression
$B = (A/\dot \gamma) \exp (-\sigma_m/\sigma_o)$.
Integrating Eq.~\ref{eq:dlnpdtheta}, using $P_s(0) = 1$,  and integrating Eq.~\ref{eq:8} by parts, we can write 
$\langle \sigma^n \rangle$ as:
\begin{equation}
\langle \sigma^n \rangle = n {\Sigma}^n \int {\theta}^{n-1} \exp[(B/p)(1 - \exp(p
  \theta))]d \theta
\end{equation}
Here we define $p = {\Sigma}/\sigma_o$.  In the special case $n=1$, it
is possible to relate this integral to known
functions\cite{abramowitz72}, but this does not appear to be true for
the general case, and it is now convenient to introduce a
dimensionless rate, $s = p/B$:
\begin{equation}
s = (\dot \gamma {\Sigma})\exp(\sigma_m/\sigma_o)/(A \sigma_o)
\end{equation}
Differentiating $\langle
\sigma^n \rangle$ with respect to $s$ yields, after a bit of algebra:
\begin{equation}
s d \langle \sigma^n \rangle /ds = -\langle \sigma^n \rangle /s +
n\sigma_o\langle \sigma^{n-1} \rangle.
\label{eq:diffeqn}
\end{equation}
To calculate the time-averaged stress $\bar \sigma$, we need to calculate $\langle \sigma \rangle$ and $\langle \sigma^{2} \rangle$.
The average  $\langle \sigma \rangle$,
\begin{equation}
d \langle \sigma \rangle /ds = -\langle \sigma \rangle /s^2 +
\sigma_o/s.
\label{eq:diffeq}
\end{equation}
We are concerned with the strongly activated regime of $A << \dot
\gamma$ and $\sigma_{m} >> \sigma_{0}$, $s >> 1$, which justifies an
expansion involving $s$ and its logs.  In lowest order when $s << 1$:
\begin{equation}
\langle \sigma \rangle \simeq \sigma_o \ln (s) + C
\end{equation}
or
\begin{equation}
\langle \sigma \rangle = \sigma_m + \sigma_o \ln ({\Sigma}/\sigma_o)
+ \sigma_o \ln (\dot \gamma /A) + C_{1},
\label{eq:rate-pred}
\end{equation}
where $C_{1}$ is a constant of integration.  Similarly, keeping the leading terms in $s$, the expression for $\langle \sigma^{2} \rangle$ is:
\begin{eqnarray}
d \langle \sigma^{2} \rangle /ds &\simeq & 2 \sigma_{0} \langle \sigma \rangle /s \nonumber \\
\langle \sigma^{2} \rangle &=&  \sigma_{0}^{2} (\ln s)^{2} + C_{2}
\label{eq:revised}
\end{eqnarray}
where $C_{2}$ is another integration constant.  Assuming the integration constants are much smaller than $\ln (s))$, 
\begin{equation}
\bar{\sigma} = {{\langle \sigma^{2} \rangle}  \over {\langle \sigma \rangle}} \simeq \sigma_{0} \ln s
\label{eq:revised2}
\end{equation}
In fact, we can now see if omitting the term $-\langle \sigma \rangle
/s^2$ at lowest order is self consistent.  We estimate $\langle \sigma
\rangle \simeq \sigma_m$, and then obtain a ratio of the two terms on
the right side of Eq.~\ref{eq:diffeq}:
\begin{equation}
[\langle \sigma \rangle/s^2]/[{\Sigma}/(s p)] \simeq
  (\sigma_m/{\Sigma})(A/\dot \gamma) \exp(-\sigma_m/\sigma_o).
\end{equation}
Indeed, this ratio should be small, so the dominant rate effect on the
mean stress should be a logarithmic stengthening.

We now turn to how the {\em mean--i.e. long-time averaged} stresses
from the  whole network within a measurement domain would be
manifested in a continuous shear experiment.  For instance, in the 2D
Couette experiments\cite{hartley_03}, time-dependent force data are
obtained in a finite region comprising roughly 10\% of the whole
system.  In 3D experiments\cite{daniels05a}, pressure measurements are
made over an area which contacts several tens of particles.  In both
cases, the mean stress is computed as a time average.  Here, we focus
on the 2D experiment, since it has yielded data over a range of
densities.  If $\phi$ were such that on average only one force
  chain existed at a time, then our analysis so far would yield the
mean stress. However, in general, it is necessary to adjust this
result for the mean number of  chains (or failure events),
$N(\phi)$, that are generated per unit time interval or angular
displacement in the region of interest.  Note that to lowest
order, this is simply a function of $\phi$.  In principle, we can
determine this quantity and hence $\sigma_o$ in order to compare to
preditions from the stress ensemble\cite{henkes05,henkes06}.  For
instance, from recent experiments by Sperl et al. on the same 2D
Couette system\cite{sperl08}, we estimate (at fixed $\dot{\gamma}$)
that $N(\phi) \propto (\phi - \phi_c)^a$ and $\sigma \propto ((\phi -
\phi_c)^b$, where $a \simeq 2$ and $b \simeq 1$. Data from Hartley et
al.\cite{hartley_03}, such as Fig.~\ref{fig:loggam}, yield $N(\phi)
\sigma_o$ by taking the slope of $\sigma/\sigma_{DC}$ vs. $\ln
(\dot{\gamma})$.  Unfortunately, the current data is not sufficiently
precise to give a good determination of $\sigma_o$ by this process.

Nevetheless, Eq.18, with the above $\phi$-dependent corrections, makes
a prediction, the key result of this work: $\bar \sigma \propto \ln
(\dot{\gamma})$, with a proportionality constant of $\sigma_o$,
multiplied by $N(\phi)$.  Qualitatively, the strengthening of the
material with shear rate occurs because the failure probability per
unit time for a given $(\sigma_{m} - \sigma)/\sigma_o$ is independent
of $\dot{\gamma}$ whereas the buildup of stress grows linearly with
$\dot{\gamma}$.  There is, therefore, a competition between the stress
buildup and the failure process.  If the failure rate were independent
of $\sigma$ (large $\sigma_{0}$ limit), then, it follows from
Eq. \eqref{linear} that the average stress would grow linearly with
the shear rate, as in Newtonian fluids.  In the opposite limit of
failure occurring only if the force chains are loaded up to
$\sigma_{m}$ ($\sigma_{0} \rightarrow 0$ limit), there would be a
fixed time $\propto 1/\dot \gamma$ between failures and the average
stress would become independent of the shear rate.  The exponential
increase of the failure rate with $\sigma$ leads to the logarithmic
strengthening.  Interestingly, Eq.~\ref{eq:diffeqn} indicates a much
weaker rate dependence for the variance, $V = \langle \sigma^2 \rangle
- (\langle \sigma \rangle)^2$: $dV/ds = (-V + (\langle \sigma
\rangle)^2)/s^2$.

\begin{figure}
\centerline{\includegraphics[width=0.8\columnwidth]{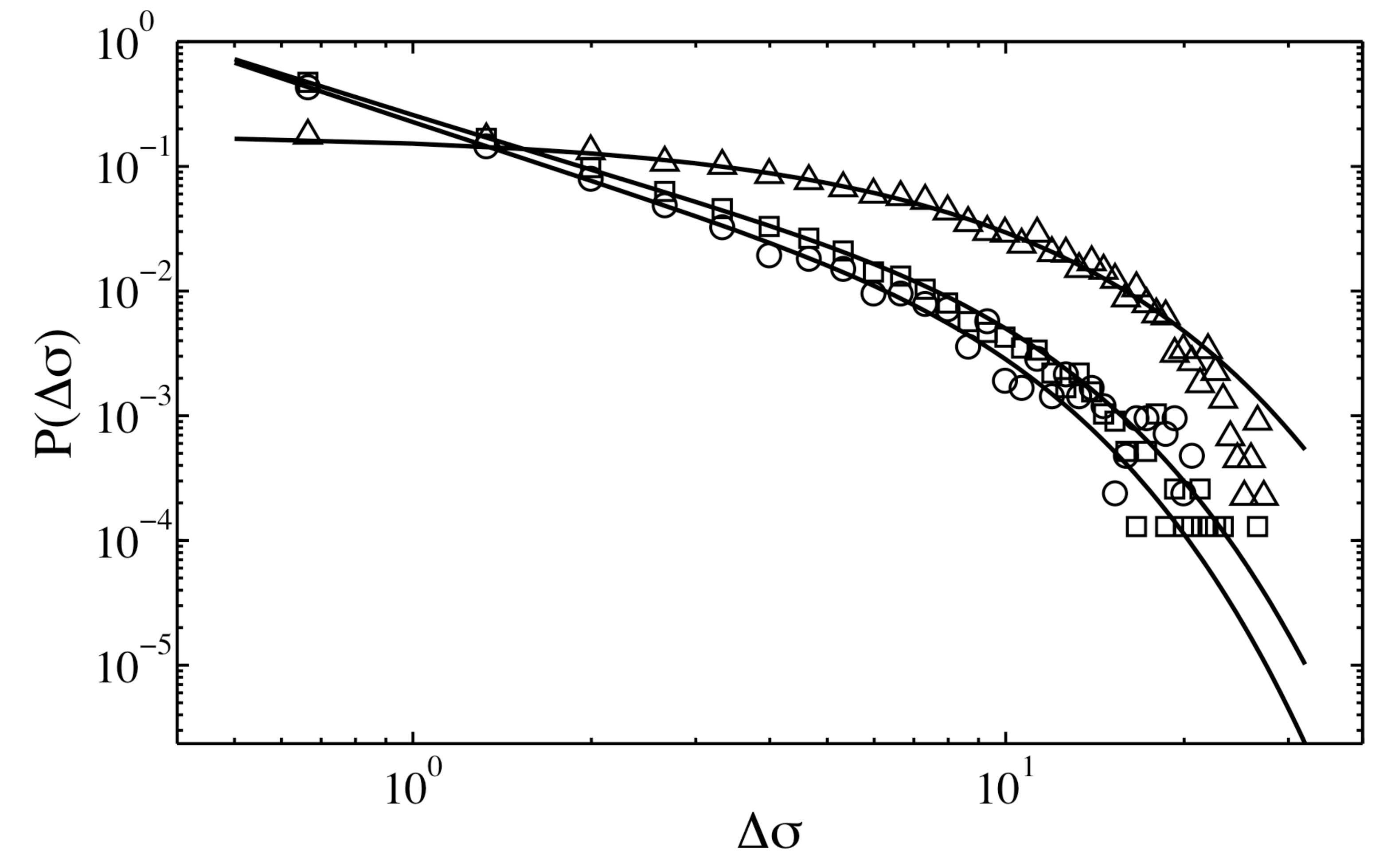}}
\caption{\label{fig:stressdrop} Fits (solid lines) to
  Eq. \ref{stressdrop} of experimental data for $\Delta \sigma$ at
  $\phi - \phi_{c} = 0.01373$, and $\dot \gamma$'s of 0.066 mHz
  (circle), 0.6645 mHz (square) and 13.307 mHz (triangle).  The fits
  yield $\sigma_{0} \simeq 2.0$, independent of $\dot \gamma$ and
  $\sigma_{m}^{0}/\sigma_{0}\simeq 1.58 ~, 1. 82 ~, 2.0 $ (lowest to
  the highest value of $\dot \gamma$.)}
\end{figure}

The distribution of the stress drops, $\Delta \sigma$, occuring for
each ``avalanche" in a time series is sensitive to the distribution of
$\sigma_{m}$, and this distribution can be calculated
exactly\cite{sollich98,henkesunpub} for an experimentally
realistic\cite{gengdrag} exponential distribution of $\sigma_{m}$:
$e^{-\sigma_{m}/\sigma_{m}^{0}}$.  For $\Sigma\dot
\gamma/\sigma_{0} << A (e^{\Delta \sigma/\sigma_{0}}- 1)$, the
distribution is
\begin{equation}
P(\Delta \sigma) = e^{\Delta \sigma /\sigma_{0}}\big{[} {1 \over {e^{\Delta
\sigma/\sigma_{0}}-1}}\big {]}^{(1+\sigma_{0}/\sigma_{m}^{0})} ~,
\label{stressdrop}
\end{equation}
which we have fitted to the experimental data for stress drops,
ensuring that the chosen data meet the criterion for which
Eq. \ref{stressdrop} is applicable.   Representative data at $\phi -
\phi_{c} = 0.01373$ are shown in Fig.~\ref{fig:stressdrop}. The ratio of the two 
fitting parameters, $\sigma_{0} / \sigma_{m}^{0}$, follows by fitting the exponent of the
power law regime, and $\sigma_{m}^{0}$ can be estimated from a fit to
the exponential tail.  We use these values to constrain the fitting
to the full form of Eq. \ref{stressdrop}.

The analysis presented in this letter is reminiscent of the soft
glassy rheology (SGR) model\cite{sollich_prl97,sollich98} with the
important distinction that the role of energy is being played by the
stress.  The noise temperature $x$ of the SGR is replaced by
$1/\alpha$ in the stress-based ensemble.  The SGR model incorporates
disorder through a distribution of activation barriers, which in the
context of the current work would translate to a distribution of
$\sigma_{m}$.  The average stress is not sensitive to the distribution
of $\sigma_{m}$ in the large $s$ limit, and scales as $(\ln {\dot
  \gamma})$.  The difference with the SGR result, $(x \ln {\dot
  \gamma})^{1/2}$, can be explained through replacement of an energy
barrier by a stress barrier\cite{henkesunpub}.

The present analysis is a first step towards understanding linear
logarithmic strengthening in granular materials which distinguishes
granular rheology, with dissipative interactions, from that of other
materials.  A key point is that a stress ensemble rather than a energy
ensemble yields the correct rate scaling.  
Force networks are visually obvious in the experiments,
but their quantitative connection to the complete stress states has not yet established.  Characterization of mesoscale structures
such as force networks, and their connection to macroscopic variables
such as stress remains a great challenge for the field.  Adopting the
framework of the SGR model with its meanfield approach to correlations
but with the noise temperature and energy replaced by their
counterparts from the stress-based ensemble should provide a fruitful
avenue for building these connections.

This work was supported by NSF-DMR0555431,  the
US-Israel Binational Science Foundation \#2004391, and, NSF-DMR0549762.

\bibliographystyle{apsrev}

\end{document}